\documentclass[useAMS,usenatbib]{mn2e}

\usepackage{lscape}
\usepackage{epsfig}
\usepackage{ltxtable}
\usepackage{array}

\title[X-ray selected AGN from the 6dFGS]{The RASS--6dFGS catalogue: a sample of X-ray selected AGN from the 6dF Galaxy Survey}
\author[Mahony et al.]{Elizabeth K. Mahony$^{1}$\thanks{E-mail:emahony@physics.usyd.edu.au}, Scott M. Croom$^{1}$, Brian J. Boyle$^{2}$, Alastair C. Edge$^{3}$, \and Tom Mauch$^{4}$ and Elaine M. Sadler$^{1}$\\
$^{1}$Sydney Institute for Astronomy, School of Physics, University of Sydney, NSW 2006, Australia\\
$^{2}$Australia Telescope National Facility, CSIRO, P.O Box 76, Epping, NSW 1710, Australia\\
$^{3}$Department of Physics, University of Durham, South Road, Durham DH1 3LE\\
$^{4}$Oxford Astrophysics, Department of Physics, Keble Road, Oxford OX1 3RH
}
\begin{document}

\date{Accepted 2009 .... Received 2009 ...; in original form 2009 ...}

\pagerange{\pageref{firstpage}--\pageref{lastpage}} \pubyear{2009}

\maketitle

\label{firstpage}

\begin{abstract}

We present a catalogue of 3405 X-ray sources from the {\it ROSAT} All Sky Survey (RASS) Bright Source Catalogue which fall within the area covered by the 6dF Galaxy Survey (6dFGS). The catalogue is count-rate limited at $0.05$\,cts\,s$^{-1}$ in the X-ray and covers the area of sky with $\delta < 0^{\circ}$ and $|b|>10^{\circ}$.  The RASS--6dFGS sample was one of the additional target catalogues of the 6dFGS and as a result we obtained optical spectra for 2224 (65$\%$) RASS sources. Of these, 1715 (77$\%$) have reliable redshifts with a median redshift of z=0.16 (excluding the Galactic sources). For the optically bright sources ($b_{\rm J} \leq 17.5$) in the observed sample, over $90\%$ have reliable redshifts. The catalogue mainly comprises QSOs and active galaxies but also includes 238 Galactic sources. Of the sources with reliable redshifts the majority are Type 1 AGN (69$\%$), while 12$\%$ are Type 2 AGN, 6$\%$ absorption-line galaxies and 13$\%$ are stars. We also identify a small number of optically-faint, very low redshift, compact objects which fall outside the general trend in the $b_{\rm J}-z$ plane. The RASS--6dFGS catalogue complements a number of northern hemisphere samples, particularly the RBSC--NVSS sample \citep{bauer}, and furthermore, in the same region of sky ($-40^{\circ}<\delta<0^{\circ}$) reveals an additional 561 sources that were not identified as part of that sample. 

We detect 918 sources (27$\%$) of the RASS--6dFGS sample in the radio using either the 1.4\,GHz NRAO VLA Sky Survey (NVSS) or the 843\,MHz Sydney University Molonglo Sky Survey (SUMSS) catalogues and find that the detection rate changes with redshift. At redshifts larger than 1 virtually all of these sources have radio counterparts and with a median flux density of 1.15\,Jy, they are much stronger than the median flux density of 28.6\,mJy for the full sample. We attribute this to the fact that the X-ray flux of these objects is being boosted by a jet component, possibly Doppler boosted, that is only present in radio-loud AGN.  

The RASS--6dFGS sample provides a large set of homogeneous optical spectra ideal for future studies of X-ray emitting AGN.

\end{abstract}


\section{Introduction} \label{intro}

Large area surveys of Active Galactic Nuclei (AGN) have become increasingly prevalent across all wavelengths over the past decade (e.g. SDSS; \citealt{SDSSDR6}, 2QZ; \citealt{2004MNRAS.349.1397C}, NVSS; \citealt{nvss}, SUMSS; \citealt{sumss}). These surveys provide a statistically significant sample of objects allowing us to probe different source populations across a wide range of redshifts. One common aim of many of these surveys is to study the evolution of active galaxies and their central SMBH. Studies have shown that the mass of the SMBH and the mass and luminosity of the host galaxy are strongly linked \citep{1995ARA&A..33..581K, 1998AJ....115.2285M}, suggesting that their evolution is also correlated. There still remains many open questions regarding not only the evolution of AGN, but also the inner workings of the central engine itself. These include whether there is a link between star formation and AGN activity, what role feedback plays in the evolution of the SMBH as well as the role of the surrounding environment.

The X-ray emission is believed to originate from near the centre of the AGN in the form of a hot corona surrounding the accretion disk \citep{1991ApJ...380L..51H}. This gives us insight into the underlying physics of accreting black holes and provides a clean signature of AGN activity. From the unified model of AGN \citep{antonucci}, there are now many diagonostics used in identifying AGN, primarily in the optical regime. For this reason, the majority of AGN catalogues have been optically selected and then followed up in other wavebands. 

With the advent of the {\it Chandra} \citep{chandra} and {\it XMM-Newton} \citep{xmm} telescopes, increased spatial resolution and sensitivity has allowed the mid to high redshift Universe to be explored \citep{2005ARA&A..43..827B}. However, there is still a great deal unknown about the low redshift X-ray sky. The {\it ROSAT} All Sky Survey (RASS) provides an ideal sample of objects to fill in this gap and provide information on the latter stages of galaxy evolution. 

The RASS is the most comprehensive survey covering the soft X-ray band (0.2 - 2\,keV). The survey contains over 100,000 sources; a mixture of active galaxies, QSOs, BL-Lac objects, galaxy clusters as well as Galactic X-ray sources such as active M-stars. The {\it ROSAT} Bright Source Catalogue (RBSC), a subset of the All Sky Survey, is count-rate limited at 0.05\,cts\,s$^{-1}$ ($\sim 7\times 10^{-13}$\,erg\,s$^{-1}$\,cm$^{-2}$) and contains 18,811 sources distributed over the entire sky \citep{rass}. 

This paper presents a catalogue of X-ray selected sources from the RBSC in the southern hemisphere. These have been followed-up spectroscopically using the 6dF spectrograph, providing a unique and highly uniform catalogue of X-ray selected AGN. Previous studies have looked at identifying the correct optical counterparts of RASS sources (e.g. \citealt{2004A&A...427..387F}, \citealt{2004ApJ...616.1284M}, \citealt{2000ApJS..131..335R}) but most have not included any redshift information. 

Our RASS--6dFGS catalogue complements similar samples in the northern hemisphere. \citet{2007AJ....133..313A} present a catalogue of RASS selected AGN with SDSS spectroscopic data and \citet{bauer} present the RBSC--NVSS sample which crossmatched the RBSC with the 1.4\,GHz NRAO VLA Sky Survey (NVSS; \citealt{nvss}) and followed up with optical spectroscopy. The latter revealed a sample of 1557 bright X-ray sources with a low average redshift of $\langle z \rangle \approx 0.1$. Radio counterparts in NVSS were used to obtain accurate positions and were then followed up spectroscopically to find redshifts. Comparisons with this survey can give an insight into the radio properties of RASS sources and reveal any bias introduced by either a radio selected sample or an optically selected sample.

In this paper we describe the object selection process and the resultant spectra in Sections 2 and 3 respectively. Section 4 details how the 6dFGS redshifts compare to those found in the literature and Section 5 describes the radio detections of RASS--6dFGS sources. The data catalogue is presented in Section 6. This is followed by a discussion of the main properties of the sample along with a comparison with other X-ray selected samples in Section 7. We summarise the main results in Section 8.

\section{Object Selection}

Objects were selected such that the dominant source of X-ray emission originates from an AGN. The target list was selected from the southern sources ($\delta \leq 0^{\circ}$) of the {\it ROSAT} Bright Source Catalogue, a total of 9578 sources. Sources were then checked for optical identifications via a visual inspection process using Digitized Sky Survey (DSS) images. The majority of the optical positions were taken from the United States Naval Observatory (USNO) database, with the remainder taken from either the Automated Plate Measuring (APM) or DSS catalogues. Positions from these latter catalogues were used when the USNO appeared to give an incorrect position according to the DSS images. Optical magnitudes were taken from the USNO--A2.0 catalogue \citep{1998AAS...19312003M}.

To further ensure the catalogue was dominated by AGN, stellar sources were removed by a combination of two methods. Firstly, stellar sources with $b_{\rm J} < $ 14 were excluded and then candidate M-stars were removed based on a colour selection using the 2MASS first data releases. When the input catalogue was compiled, at the beginning of the 6dFGS, just less than half of the BSC sources had 2MASS information and as a result only $\sim 50\%$ of stars remaining in the X-ray sample were removed in this process. Candidate clusters of galaxies were also excluded from the target list, selected based on the X-ray extent and apparent clustering in optical images. Since the majority of the X-ray emission originates from the intracluster medium for these sources, the number of AGN removed will be relatively small \citep{2007ApJ...664..761M, 2006ApJ...644..116M}. A number of these cluster galaxies have also been observed as part of the primary 6dFGS and are discussed in Section \ref{clusters}. 

This results in a final target sample of 3405 sources which were listed as additional targets in the 6dF Galaxy Survey. Being an additional target sample meant that sources were only observed on spare fibres and hence only 65$\%$ have 6dFGS spectroscopic data. The results from this subset of sources are expected to be typical of the entire population. 

\section{6dFGS Spectra} \label{6dfspectra}

The 6-degree Field Galaxy Survey (6dFGS) is a spectroscopic survey containing redshifts and optical spectra of 136,304 sources with $\delta <$ 0$^\circ$ and Galactic latitude $|b|>$ 10$^\circ$. The survey was carried out on the UK Schmidt Telescope located at Siding Spring Observatory. Central to the survey is the 6dF Multi-Fibre Spectrograph which uses 6.7$''$ optical fibres and is capable of recording 150 spectra simultaneously over a 5.7$^\circ$ diameter field. An adaptive tiling algorithm was used to distribute these fields in such a way that maximised uniformity and completeness across the entire southern sky, covering a total area of 17,046 deg$^2$. 6dFGS spectra have a resolution of $R \sim 1000$ over the wavelength range 4000-7500\,\AA. \citet{6df, 6dfDR2, Jones} give a complete description of the survey. The final 6dFGS data release contains 179,263 targets and 117,405 unique and reliable redshifts with $cz>$ 600\,km\,s$^{-1}$. It is available online at http://www-wfau.roe.ac.uk/6dFGS/dr3/ \citep{6df2009}.

The primary list of targets was selected from the 2MASS Extended Source Catalogue (2MASS XSC; \citealt{2mass}) with K$\leq$12.5. The remaining targets come from a number of smaller additional target samples. Selection criteria for these sources are detailed in \citet{6df}. For the purpose of this paper, we are interested in those selected from the {\it ROSAT} All Sky Survey \citep{rass}.

Redshifts were measured from the 6dFGS spectra using the RUNZ package (originally written by W. Sutherland, \citealt{2001MNRAS.328.1039C}) and assigned a quality, $Q$, based on the reliability of the redshift. A quality $Q=$ 1 or 2 denotes a unreliable redshift, $Q=$ 3 signifies a probable redshift, and $Q=$ 4 implies a certain redshift. Galactic sources were assigned a quality $Q=$ 6 \citep{6df}. During the redshifting process, unusual spectra or features were flagged with comments. These comments appear in the final catalogue and are explained in more detail in Section \ref{comments}.    

\begin{table}
\begin{center}
\caption{Redshift qualities of RASS sources. Qualities $Q\geq3$ are regarded as reliable and $Q=0$ means that the source was not observed as part of the 6dFGS. Percentages are given with respect to the full catalogue in column 3 (3405 sources) and with respect to only the sources in the observed sample in column 4 (2224 sources). }
\label{qualitytab}
\begin{tabular}{cccc}
\hline
\bf Q	& \bf No. & \bf $\%$  & \bf $\%$ \\	
& & Full & obs. \\
\hline
0 & 1181 & 34.7 & -- \\
1 & 412 & 12.1 & 18.5 \\  
2 & 97 & 2.8 & 4.4 \\
3 & 123 & 3.6 & 5.5 \\
4 & 1354 & 39.8 & 60.9 \\
6 & 238 & 7.0 & 10.7 \\
\hline
\end{tabular}
\end{center}
\end{table}

\begin{figure}
  \centerline{\epsfig{file=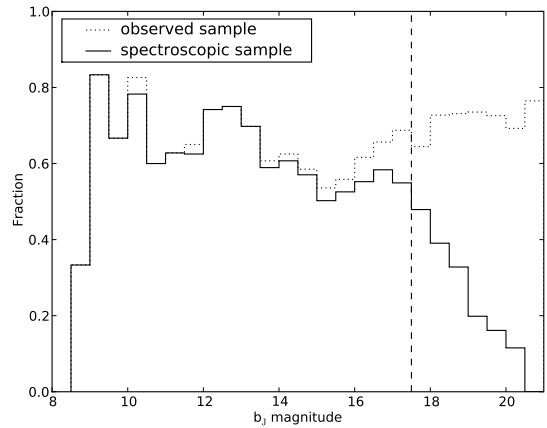, width=\linewidth}}
  \caption{Completeness of the entire sample as a function of optical magnitude. The dotted line denotes the fraction of the catalogue observed as part of the 6dFGS while the solid line shows the fraction of those that have a reliable redshift. There is one source with $b_{\rm J}$=21.5 which is not plotted due to poor statistics in that bin. $90\%$ of sources brighter than $b_{\rm J}=17.5$ have reliable redshifts, marked by the dashed line.}
  \label{completenessbmagfig}
\end{figure}

There are a total of 3405 RASS sources used as input targets for the 6dF Galaxy Survey, here defined as the RASS--6dFGS `full catalogue'. Of those sources, 2224 (65.3$\%$) were observed as part of 6dFGS and form the `observed sample'. Investigating the quality of each redshift (Table \ref{qualitytab}), revealed that there are a total of 1715 (77.1$\%$) RASS--6dFGS sources with reliable redshifts ($Q\geq3$). This subset forms the `spectroscopic sample'. Figure \ref{completenessbmagfig} shows the fraction of sources observed, binned by $b_{\rm J}$ magnitude, along with the fraction of sources with reliable redshifts. Of the 1478 optically bright objects ($b_{\rm J} \leq 17.5$) in the observed sample, 1333 (90.2$\%$) have reliable redshift measurements (marked by the dashed line in Figure \ref{completenessbmagfig}). As expected, the number of sources in the spectroscopic sample decreases as the objects become optically fainter. The selection of observed targets were not dependant on optical magnitude, but assigned according to where there were spare fibres on the 6dF spectrograph. 

\begin{figure*} 
  \centerline{\epsfig{file=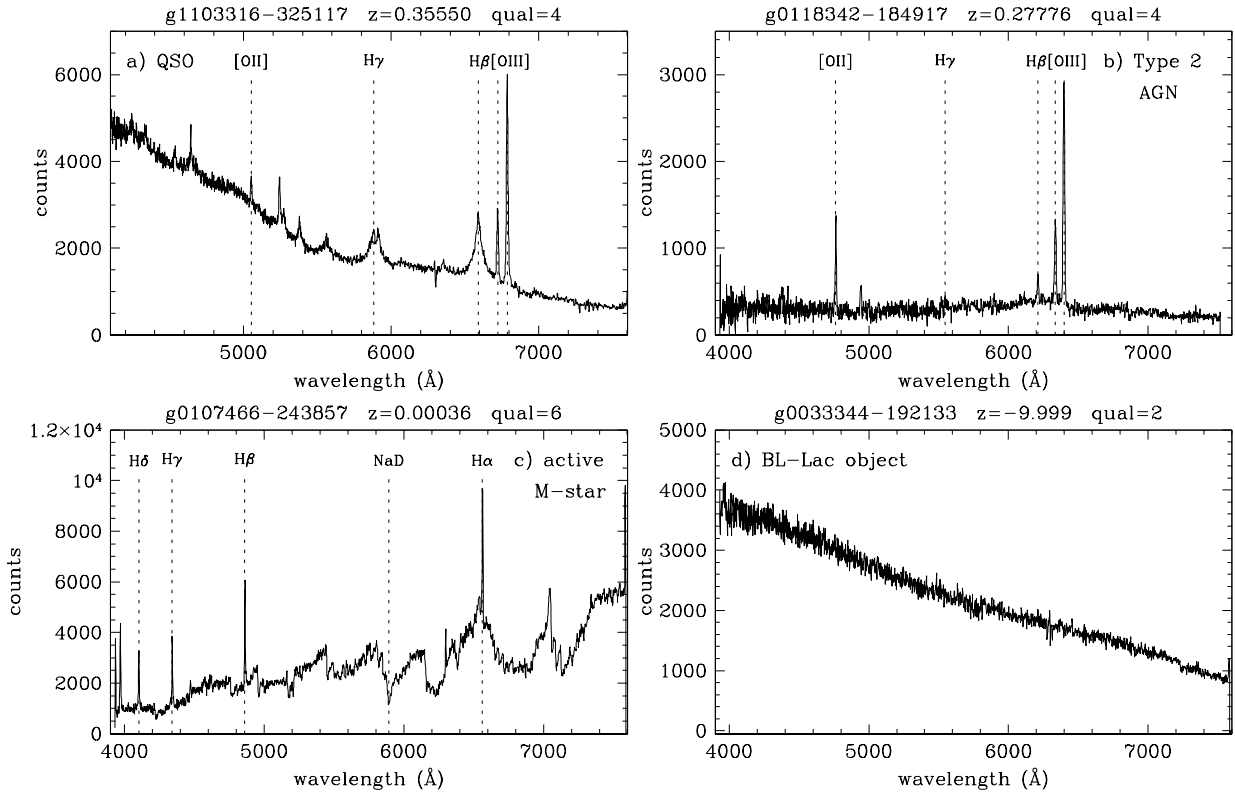, width=\linewidth}}
\caption{Example 6dFGS optical spectra. The dashed lines mark where features would appear at a given redshift, but these features may or may not be evident.}
\label{spectrafig}
\end{figure*}

There are 108 objects that were observed twice by 6dF and 8 objects observed three times. A decision was made based on the quality of the redshifts as to which observation appears in the final catalogue such that each object is only listed once. If both observations resulted in poor quality redshifts the first catalogue line was chosen, otherwise the observation with the highest quality appears in the final catalogue. In all cases but one, sources with more than one good quality spectrum had identical redshifts so once again the first line was chosen. g1446036-025346 was observed on 2 occasions and has redshifts of $z=0.0772$, $Q=3$ and $z=0.0000$, $Q=6$ listed for each observation. The spectrum suggests that this source is an active M-star so the $z=0$ redshift was included in the final catalogue. Some example 6dFGS spectra are shown in Figure \ref{spectrafig}. 

\subsection{Galactic X-ray Sources} \label{galacticsources}

Of the sources in the observed sample, there are 238 Galactic sources (10.6$\%$). The majority of these are either typical F or G type stars ($\sim 45\%$), implying either a misidentification or a foreground star dominating the spectrum, or active M-stars ($\sim 30\%$, See Figure \ref{spectrafig}). There are also a number of cataclysmic variable (CV) stars displaying strong Balmer emission and white dwarfs (listed with comments `WD' in the final data catalogue). A small number of objects have $z=0$ nebula emission, characterised by typical nebula emission lines such as the forbidden [OIII] $\lambda 5007$ line. This is generally due to foreground nebula emission dominating the spectrum. Due to the 6dFGS survey limit all of these sources have $|b|>10^{\circ}$, but 55$\%$ of these objects have $10^{\circ}<|b|<30^{\circ}$.

\subsection{Possible BL-Lac Objects} \label{bllacs}

In addition to sources marked as possible BL-Lacs during the redshifting process, spectra with poor quality redshifts (i.e. sources in the observed sample, but not in the spectroscopic sample) were checked individually by eye to search for possible BL-Lac objects. These were identified by a featureless optical spectrum with a strong continuum. An example spectrum is shown in Figure \ref{spectrafig}. Only sources with $b_{\rm J} \leq 17.5$ were inspected as beyond this magnitude it is hard to distinguish if the spectrum is featureless because it is a BL-Lac object or because of low signal to noise. There are a total of 55 (3.7$\%$) sources flagged as possible BL-Lacs, but further observational follow up is required to confirm this classification. There are 31 sources that have reliable redshifts and yet have still been flagged as possible BL-Lacs. This is because weak features were able to be distinguished (generally narrow ionised oxygen emission lines), but the spectrum was still typical of a BL-Lac object. Of these 86 sources, 42 (49$\%$) also have radio counterparts in either the SUMSS or NVSS catalogues (see Section \ref{radiodetections} for more details on how these were identified). Possible BL-Lac objects are flagged with a comment ``BLLac" or ``BLLac?" in the final catalogue, where the question mark denotes a higher degree of uncertainty in the classification.

\section{Redshifts in the Literature} \label{NEDredshifts}

To check for previously known redshifts in the RASS--6dFGS full sample we searched the NASA Extragalactic Database (NED). Excluding redshifts listed with a 6dFGS reference (from previous data releases), it was found that 995 (29.2$\%$) sources already had known redshifts, of which 496 had a good quality ($Q\geq3$) 6dFGS redshift. Of the 995 sources with known redshifts, 826 (83$\%$) sources are optically bright ($b_{\rm J} \leq 17.5$) and of these bright sources 409 have a reliable 6dFGS redshift. The SIMBAD database was also checked, revealing 8 stars with existing classifications. This smaller sample of 504 (496 from NED and 8 from SIMBAD) sources with both 6dFGS redshifts and a redshift from an independent source allowed us to check the reliability of the 6dFGS redshifts. 

\begin{figure}
  \centerline{\epsfig{file=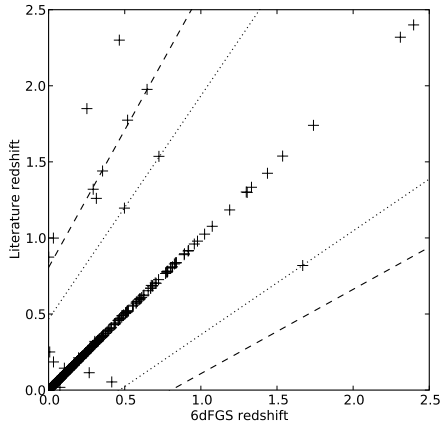, width=\linewidth}}
  \caption {6dFGS redshifts plotted against corresponding redshifts found in the literature. The dashed lines denote an incorrect identification of the CIV and Mg II emission lines, and the dotted lines represent misidentifications of CIII] and MgII. There are 18 sources with inconsistent redshifts ($\Delta z > 0.02$) which are listed in Table \ref{wrongredshiftstab}. } 
  \label{6dflitzfig}
\end{figure}

\begin{table*} 
  \begin{center}
    \begin{minipage}{\linewidth}
\caption{6dFGS sources with inconsistent redshifts in the literature. A $^{+}$ denotes references where the optical spectra were available either published within the referenced paper or online. In some cases the spectrum was not available but the observed features were listed within the publication. Sources where the literature redshift was designated as more reliable are listed in Section \ref{individsources} with explanations. Objects where we believe the 6dFGS spectrum is more reliable are shown in Appendix \ref{6dfspectraapp}. } 
    \begin{tabular}{lccccccc}
\hline
\bf 6dFGS name	& \bf z & \bf Q & \bf Observed lines (6dFGS) & \bf Lit z &\bf Observed lines (lit)& \bf Reference \footnote{B2000: \citet{bauer}, J1984: \citet{1984ApJ...286..498J}, D1987: \citet{1987PhDT........11D}, W2000:  \citet{2000A&A...358...77W}, S2004: \citet{2004AA...423..121S}, M1993: \citet{1993RMxAA..25...51M}, S1993: \citet{1993ApJ...412..541S}, S1991: \citet{1991ApJS...76..813S}, SDSS DR2: \citet{SDSSDR2}, HB1989: \citet{1989QSO...C......0H}, H2003: \citet{2003AJ....125..572H}, F1984: \citet{1984MNRAS.210...69F}, B1994: \citet{1994PhDT.......222B}, P2001: \citet{2001AA...369..787P}.}
& \bf Verdict\\	
\hline
g0023394-175354 & 0.4148 & 3 & H$\beta$, [OII] & 0.0535 &  & B2000 & Lit z \\
g0038147-245902 & 0.4981 & 4 & MgII, [OII] & 1.1960 & CIV, HeII, CIII] & J1984 & 6dFGS \\
g0140044-153256 & 1.6688 & 3 & CIV, CIII] & 0.8188 & & B2000 & 6dFGS \\
g0154019-403152 & 0.2516 & 4 & [OII], H$\gamma$, H$\beta$, [OIII]	& 1.8500 & & D1987 & 6dFGS \\
g0210462-510102 & 0.0316 & 4 & CaII K$\&$H, Mg & 0.9990 &  & W2000 & Lit z\\ 
g0250553-361636 & 0.7246 & 3 & MgII	& 1.5362 & H$\beta$, [OIII] $^{+}$ & S2004 & Lit z\\
g0301558-433051 & 0.4645 & 4 & MgII, H$\gamma$, H$\beta$ & 2.3000 & & M1993 & 6dFGS \\
g0349232-115927 & 0.0321 & 4 & H$\gamma$, H$\beta$, [OIII] & 0.1850 & CaII H$\&$K, G, Mg, H$\beta$, H$\alpha$ $^{+}$  & S1993 & Lit z\\
g0845102-073205 & 0.1036 & 4 & [OII], H$\gamma$, H$\beta$, [OIII], H$\alpha$ & 0.1440 & [OII], H$\beta$, [OIII] & S1991 & 6dFGS\\
g1026586-174859 & 0.2668 & 3 & CaII K$\&$H, G & 0.1142 & & B2000 & 6dFGS \\
g1150439-002354 & 0.6467 & 4 & MgII	& 1.9762 & CIV, HeII, CIII], MgII $^{+}$ & SDSS DR2 & Lit z \\
g1238074-115925 & 0.2930 & 4 & [OII], H$\gamma$, H$\beta$, [OIII] & 1.3200 & & HB1989 & 6dFGS \\
g1923325-210432 & 0.0008 & 6 & CaII K$\&$H, G, Mg, Na	& 0.8740 & CIII], MgII $^{+}$& H2003 & Lit z\\
g2125407-490938 & 0.0744 & 4 & CaII K$\&$H, H$\beta$, [OIII], H$\alpha$ & 0.0180 & & F1984 & 6dFGS \\
g2125521-233812 & 0.5188 & 3 & MgII & 1.7740 & Ly$\alpha$, CIV, CIII], MgII $^{+}$ & B1994 & Lit z\\
g2223114-423212 & 0.3137 & 3 & H$\beta$ & 1.2600 & & D1987 & 6dFGS\\
g2238127-394019 & 0.0074 & 4 & H$\beta$, [OIII], H$\alpha$ & 0.2506 & & P2001 & Lit z\\
g2247333-405718 & 0.3544 & 3 & [OIII] & 1.4400 & & D1987 & 6dFGS\\ 
\hline
    \end{tabular}
\label{wrongredshiftstab}
   \end{minipage}
  \end{center}
\end{table*}

Plotting 6dFGS redshifts against the corresponding redshifts found in the literature (Figure \ref{6dflitzfig}) revealed that 18 sources have significantly different redshifts with $\Delta z > 0.02$. These are listed in Table \ref{wrongredshiftstab}. Where possible the original spectrum in the literature was inspected to check the reliability of the redshift. Where the 6dFGS redshift is the less reliable of the two, the redshift was adjusted in the final catalogue. Reasons behind the incorrect 6dFGS redshifts are given in Section \ref{individsources}. This method unearthed only 8 (1.6$\%$) out of the 504 sources that had independent redshifts, where the 6dFGS spectrum was judged to be incorrect. This is slightly higher than the error rate for the primary 6dFGS survey ($1.2\%$; \citet{6df2009}) most likely due to the fact that the RASS--6dFGS sources are on average optically fainter than the primary sample. 

A small amount of NED information has been included in the main data catalogue; redshift, corresponding reference and classification (generally either `G' or `QSO').

\subsection{Notes on Individual Sources} \label{individsources}

There are a small number of sources where the redshift in the literature is more reliable than the 6dFGS redshift. Reasons for each object are as follows:

\noindent{\it g0023394-175354:} Emission line marked as [OII] in the 6dFGS spectrum actually corresponds to [OIII], which agrees with the literature redshift.\\
{\it g0210462-510102:} A similar redshift of z=1.003 \citep{1976ApJ...207L...5P} showing CIII], MgII emission lines for this object suggests the literature redshift is more reliable. Also listed in NED as a possible BL-Lac object. \\
{\it g0250553-361636:} Literature redshift is from VLT/ISAAC spectrum. Observed emission line in 6dFGS corresponds to CIII].\\ 
{\it g0349232-115927:} 6dFGS instrumentational error. Fibre cross-talk from a source being observed by an adjacent fibre. See Section 3.1 in \citet{6dfDR2}.\\
{\it g1150439-002354:} The SDSS spectrum of this object has a higher S/N ratio than the 6dFGS spectrum. Emission line observed by 6dFGS corresponds to CIV and there is also evidence of CIII].\\
{\it g1923325-210432:} Nearby star dominates the spectrum. The actual X-ray source is a QSO with redshift z = 0.874.\\  
{\it g2125521-233812:} The spectrum found in \citet{1994PhDT.......222B} displays more emission lines and has higher S/N. \\
{\it g2238127-394019:} Also fibre cross-talk.

\vspace{-0.4cm}

\section{Radio detections of RASS-6dFGS sources} \label{radiodetections}

Radio source identifications of objects in the 6dFGS-RASS catalogue were found by crossmatching it with the 1.4\,GHz NRAO VLA Sky Survey (NVSS; \citet{nvss}) at $\delta > -40^{\circ}$ and the 843\,MHz Sydney University Molonglo Sky Survey (SUMSS; \citet{sumss}) at $\delta < -30^\circ$. Candidate SUMSS and NVSS matches to 6dFGS objects were confirmed using the method outlined by \citet{mauch} for the 6dFGS--NVSS sample. The SUMSS and NVSS radio source catalogues were searched for components within 180\,arcsec of each 6dFGS position. Those with a single radio component that were separated by more than 40\,arcsec from 
\begin{landscape}
\pagestyle{empty} 
\topmargin=+1.5cm 

\begin{table} 
\scriptsize{
 \begin{center}
    \begin{minipage}{\linewidth}
\caption{The first 50 entries of the 6dFGS-RASS AGN sample. The full catalogue is available in the electronic version of the journal.
 \label{examplecat}}
    \newcolumntype{V}{>{\centering\arraybackslash} m{0.6cm} }
    \newcolumntype{W}{>{\centering\arraybackslash} m{0.1cm} }   
    \begin{tabular}{{lcccccWVVVVVVVcccccccl}}
\hline
\bf 6dFGSname & \multicolumn{2}{c}{\bf Optical Postion} & \bf $b_{\rm J}$ & \bf R & \bf RASSname & \bf E & \bf cnts/s & \bf $\pm$ & \bf N(H) & \bf Flux & \bf $\pm$ & \bf NVSS & \bf SUMSS & \multicolumn{2}{c}{\bf 6dFGS} & \bf Q & \bf ID & \multicolumn{3}{c}{\bf NED information} & \bf 6dFGS \\
& \multicolumn{2}{c}{(J2000)} & \multicolumn{2}{c}{(Mag)} & & & \tiny{$\times 10^{-2}$} & \tiny{$\times 10^{-2}$} & \tiny{$\times 10^{20}$} & \tiny{$\times 10^{-13}$} & \tiny{$\times 10^{-13}$} & Flux & Flux & \bf z & error & & & Class & z & Ref \footnote{2dFb: \citet{2003astro.ph..6581C}, S00a: \citet{2000AN....321....1S}, C97c: \citet{1997AJ....114.2353C}, D88: \citet{1988AJ.....95..284D}, W00: \citet{2000A&A...358...77W}, S89: \citet{1989A&AS...80..103S}, B99: \citet{1999A&A...347...47B}, V99: \citet{1999ApJ...525..995V}, SDSSe: \citet{2002AJ....123..485S}, R98: \citet{1998MNRAS.300..417R}, B00b: \citet{bauer}, M89: \citet{1989ApJS...69..353M}, SDSS2: \citet{SDSSDR2}, D04: \citet{2004A&A...416..515D}.} & \bf Comments\\
& Ra & Dec & & & & & &  & \tiny{(cm$^{-2}$)} & \multicolumn{2}{c}{\tiny{(erg\,cm$^{-2}$s$^{-1}$)}} & {\tiny(mJy)} & {\tiny(mJy)} & & & & & &  & & \\
\hline 
g0000097-633538 & 00 00 09.67 & -63 35 37.5 & 16.8 & 16.7 & R0000100-633543 & . & 19 & 3.12 & 0.593 & 18.9 & 3.1 & --- & --- & 0.1372 & 0.00010 & 4 & 113 & --- & --- & --- & narrow\\
g0000207-261025 & 00 00 20.72 & -26 10 24.7 & 20.2 & 17.9 & R0000195-261032 & . & 11.8 & 2.22 & 1.51 & 16.5 & 3.1 & --- & --- & -9.999 & --- & 1 & 113 & --- & --- & --- & ---\\
g0000439-260522 & 00 00 43.92 & -26 05 21.8 & 15.2 & 14.7 & R0000442-260521 & . & 11.5 & 2.23 & 1.59 & 16.3 & 3.17 & --- & --- & --- & --- & 0 & 0 & G & 0.0590 & 2dFb & ---\\
g0001171-315044 & 00 01 17.06 & -31 50 44.1 & 19.3 & 17.9 & R0001163-315041 & . & 5.67 & 1.74 & 1.21 & 7.29 & 2.24 & 7.0 & --- & -9.999 & --- & 2 & 113 & --- & --- & --- & ---\\
g0001286-320842 & 00 01 28.57 & -32 08 42.4 & 17.3 & 16.7 & R0001297-320837 & . & 13.7 & 2.56 & 1.19 & 17.5 & 3.27 & --- & --- & 0.4676 & 0.00049 & 4 & 113 & --- & --- & --- & broad\\
g0002021-103023 & 00 02 02.09 & -10 30 22.8 & 16.9 & 16.5 & R0002025-103030 & . & 6.1 & 1.62 & 2.86 & 10.7 & 2.86 & --- & --- & 0.1025 & 0.00025 & 4 & 113 & --- & --- & --- & narrow\\
g0002152-672653 & 00 02 15.18 & -67 26 53.3 & 18.5 & 17.9 & R0002175-672713 & . & 9.2 & 2.03 & 64.4 & 50.7 & 11.2 & --- & 21.4 & -9.999 & --- & 1 & 113 & --- & --- & --- & ---\\
g0002536-260347 & 00 02 53.61 & -26 03 46.7 & 17.2 & 17.1 & R0002545-260339 & . & 13.3 & 2.33 & 1.58 & 18.9 & 3.31 & --- & --- & -9.999 & --- & 2 & 113 & --- & --- & --- & BLLac?\\
g0003078-180550 & 00 03 07.86 & -18 05 49.9 & 11.8 & 12.0 & R0003076-180550 & . & 31.9 & 3.43 & 2.24 & 51.3 & 5.53 & --- & --- & 0.0540 & 0.00020 & 4 & 1 & QSO & 0.0543 & S00a & narrow\\
g0003168-275627 & 00 03 16.83 & -27 56 27.0 & 16.2 & 15.2 & R0003168-275627 & . & 11.3 & 2.19 & 1.52 & 15.8 & 3.06 & --- & --- & --- & --- & 0 & 0 & G & 0.1023 & 2dFb & ---\\
g0003196-524727 & 00 03 19.58 & -52 47 26.9 & 18.4 & 17.9 & R0003204-524727 & . & 30.7 & 5.07 & 1.52 & 43 & 7.1 & --- & 65.3 & --- & --- & 0 & 0 & --- & --- & --- & ---\\
g0003261-193331 & 00 03 26.09 & -19 33 30.6 & 17.8 & 17.5 & R0003256-193322 & . & 5.06 & 1.5 & 2.11 & 7.99 & 2.37 & --- & --- & 0.3931 & 0.00031 & 4 & 113 & --- & --- & --- & broad\\
g0004158-642525 & 00 04 16.23 & -64 25 28.7 & 17.6 & 16.8 & R0004195-642517 & . & 8.13 & 1.98 & 1.45 & 11.2 & 2.71 & --- & --- & 0.1185 & 0.00042 & 4 & 1 & --- & --- & --- & abs\\
g0004167-290235 & 00 04 16.65 & -29 02 34.7 & 18.5 & 17.9 & R0004168-290240 & . & 10 & 2.01 & 1.45 & 13.8 & 2.76 & 15.1 & --- & 0.5584 & 0.00195 & 4 & 113 & --- & --- & --- & broad\\
g0005116-300634 & 00 05 11.64 & -30 06 34.4 & 18.7 & 17.9 & R0005122-300648 & . & 7.66 & 1.92 & 1.31 & 10.2 & 2.54 & --- & --- & 0.2060 & 0.00075 & 4 & 113 & --- & --- & --- & broad\\
g0005204-742640 & 00 05 20.44 & -74 26 40.2 & 15.9 & 15.4 & R0005212-742632 & . & 36.1 & 3.17 & 4.91 & 77.6 & 6.82 & --- & --- & --- & --- & 0 & 0 & G & 0.1316 & C97c & ---\\
g0005370-214103 & 00 05 37.00 & -21 41 03.4 & 17.8 & 16.5 & R0005370-214111 & . & 12.9 & 2.32 & 1.87 & 19.4 & 3.5 & --- & --- & 0.2574 & 0.00058 & 4 & 113 & --- & --- & --- & broad\\
g0005431-500655 & 00 05 43.06 & -50 06 54.7 & 12.9 & 12.2 & R0005431-500703 & . & 37.9 & 6.96 & 3.13 & 69.1 & 12.7 & --- & --- & --- & --- & 0 & 0 & G & 0.0335 & D88 & ---\\
g0005447-080824 & 00 05 44.74 & -08 08 23.8 & 17.9 & 18.6 & R0005451-080822 & . & 6.38 & 1.54 & 3.1 & 11.6 & 2.8 & --- & --- & 0.1754 & 0.00010 & 4 & 113 & --- & --- & --- & broad\\
g0005585-275858 & 00 05 58.54 & -27 58 57.6 & 17.6 & 17.4 & R0005582-275854 & . & 6.5 & 1.68 & 1.63 & 9.31 & 2.41 & 308.2 & --- & 0.6264 & 0.00154 & 4 & 113 & QSO & 0.6250 & W00 & broad\\
g0006139-062335 & 00 06 13.88 & -06 23 35.3 & 19.1 & 17.8 & R0006135-062347 & E & 6.97 & 1.63 & 3.1 & 12.7 & 2.97 & 2050.7 & --- & 0.3468 & 0.00061 & 4 & 113 & G & 0.3470 & S89 & narrow+BLLac\\
g0006145-291221 & 00 06 14.45 & -29 12 20.8 & 18.0 & 17.1 & R0006151-291223 & . & 12.1 & 2.25 & 1.47 & 16.7 & 3.12 & --- & --- & 0.3974 & 0.00042 & 4 & 113 & --- & --- & --- & broad\\
g0006263-405120 & 00 06 26.31 & -40 51 20.3 & 16.3 & 16.8 & R0006269-405136 & . & 16 & 3.22 & 5.28 & 35.3 & 7.1 & --- & --- & 0.1605 & 0.00011 & 4 & 113 & --- & --- & --- & broad\\
g0006469-332310 & 00 06 46.89 & -33 23 10.2 & 17.5 & 17.4 & R0006472-332316 & . & 14.1 & 2.45 & 1.14 & 17.8 & 3.09 & 7.4 & --- & 0.2170 & 0.00009 & 4 & 113 & G & 0.2170 & B99 & broad+bad$\_$splicing\\
g0006526-070715 & 00 06 52.60 & -07 07 14.7 & 18.0 & 16.8 & R0006541-070721 & . & 5.79 & 1.62 & 3.06 & 10.5 & 2.93 & --- & --- & -9.999 & --- & 1 & 113 & --- & --- & --- & ---\\
g0007292-450222 & 00 07 29.18 & -45 02 22.0 & 17.0 & 16.8 & R0007305-450213 & . & 9.62 & 2.6 & 3.1 & 17.5 & 4.72 & --- & --- & 0.3574 & 0.00029 & 3 & 113 & QSO & 0.3600 & V99 & narrow+BLLac?\\
g0007410-635146 & 00 07 41.01 & -63 51 46.0 & 17.0 & 16.7 & R0007408-635152 & . & 7.08 & 1.88 & 5.58 & 15.9 & 4.23 & --- & --- & 0.5584 & 0.00082 & 4 & 113 & --- & --- & --- & broad\\
g0007432-695948 & 00 07 43.16 & -69 59 47.8 & 18.3 & 17.6 & R0007440-695956 & . & 5.71 & 1.42 & 57.5 & 30.2 & 7.53 & --- & --- & -9.999 & --- & 1 & 113 & --- & --- & --- & ---\\
g0007570-115145 & 00 07 56.97 & -11 51 45.1 & 17.9 & 16.0 & R0007585-115145 & . & 6.05 & 1.63 & 2.64 & 10.4 & 2.78 & --- & --- & 0.2072 & 0.00044 & 4 & 113 & --- & --- & --- & abs\\
g0008026-030948 & 00 08 02.60 & -03 09 47.8 & 18.8 & 17.5 & R0008032-030942 & . & 8.16 & 1.78 & 3.42 & 15.4 & 3.35 & --- & --- & 0.3677 & 0.00092 & 3 & 113 & --- & --- & --- & narrow\\
g0008133-005753 & 00 08 13.25 & -00 57 53.3 & 17.7 & 16.0 & R0008133-005752 & . & 5.08 & 1.36 & 3.23 & 9.36 & 2.5 & --- & --- & 0.1391 & 0.00010 & 4 & 113 & G & 0.1390 & SDSSe & broad\\
g0008271-405127 & 00 08 27.07 & -40 51 27.2 & 11.5 & 14.0 & R0008272-405137 & . & 8.99 & 2.37 & 4.99 & 19.4 & 5.12 & --- & --- & 0.0614 & 0.00009 & 4 & 8 & G & 0.0617 & R98 & narrow\\
g0008306-155817 & 00 08 30.59 & -15 58 17.4 & 18.1 & 17.4 & R0008305-155825 & . & 14.9 & 2.35 & 2.02 & 23.1 & 3.64 & 6.1 & --- & 0.2945 & 0.00078 & 3 & 113 & --- & --- & --- & narrow\\
g0008354-233928 & 00 08 35.41 & -23 39 28.0 & 17.4 & 16.7 & R0008354-233917 & . & 23.3 & 2.89 & 2.32 & 38.1 & 4.71 & 35.8 & --- & -9.999 & --- & 2 & 113 & G & 0.1470 & B00b & BLLac\\
g0008461-464911 & 00 08 46.10 & -46 49 10.6 & 14.9 & 14.6 & R0008466-464920 & . & 12 & 3.04 & 2.32 & 19.6 & 4.97 & --- & --- & 0.0708 & 0.00003 & 4 & 113 & G & 0.0708 & M89 & broad\\
g0009003-014453 & 00 09 00.33 & -01 44 52.9 & 13.7 & 12.2 & R0009002-014442 & . & 17.5 & 2.32 & 3.39 & 32.9 & 4.35 & --- & --- & --- & --- & 0 & 0 & --- & --- & --- & ---\\
g0009046-103429 & 00 09 04.55 & -10 34 28.7 & 17.3 & 17.7 & R0009047-103432 & . & 8.55 & 2.01 & 2.92 & 15.2 & 3.57 & --- & --- & 0.2407 & 0.00020 & 4 & 113 & QSO & 0.2408 & SDSS2 & broad\\
g0009089-231144 & 00 09 08.94 & -23 11 43.5 & 18.6 & 17.3 & R0009096-231137 & . & 5.76 & 1.67 & 2.28 & 9.34 & 2.71 & --- & --- & -9.999 & --- & 1 & 113 & --- & --- & --- & ---\\
g0009381-344323 & 00 09 38.10 & -34 43 23.0 & 17.8 & 17.6 & R0009380-344325 & . & 8.95 & 2.08 & 1.19 & 11.4 & 2.65 & --- & --- & 0.2356 & 0.00003 & 4 & 113 & --- & --- & --- & broad\\
g0009498-431651 & 00 09 49.80 & -43 16 50.5 & 18.8 & 17.7 & R0009507-431653 & . & 12.8 & 3.09 & 1.19 & 16.4 & 3.95 & --- & 51.0 & -9.999 & --- & 2 & 113 & --- & --- & --- & ---\\
g0010049-115247 & 00 10 04.89 & -11 52 47.3 & 19.0 & 19.0 & R0010046-115254 & . & 5.56 & 1.6 & 2.5 & 9.34 & 2.69 & --- & --- & -9.999 & --- & 1 & 113 & --- & --- & --- & ---\\
g0010100-044238 & 00 10 09.98 & -04 42 37.5 & 14.5 & 14.6 & R0010102-044225 & . & 22.1 & 2.79 & 3.26 & 40.9 & 5.16 & 6.4 & --- & 0.0295 & 0.00001 & 4 & 1 & G & 0.0295 & D04 & narrow\\
g0010160-153232 & 00 10 15.97 & -15 32 32.3 & 17.3 & 16.4 & R0010165-153215 & . & 7.15 & 1.72 & 2.01 & 11.1 & 2.66 & --- & --- & 0.1815 & 0.00046 & 4 & 113 & G & 0.1805 & 2dFb & abs\\
g0010200-061706 & 00 10 19.98 & -06 17 05.8 & 15.6 & 13.7 & R0010205-061703 & . & 8 & 1.65 & 3.1 & 14.5 & 2.99 & --- & --- & 0.0820 & 0.00038 & 4 & 1 & --- & --- & --- & broad\\
g0010412-314125 & 00 10 41.21 & -31 41 25.4 & 18.8 & 17.9 & R0010415-314128 & . & 6.02 & 1.54 & 1.44 & 8.25 & 2.1 & --- & --- & 0.1815 & 0.00020 & 4 & 113 & --- & --- & --- & broad\\
g0010496-081722 & 00 10 49.60 & -08 17 22.2 & 16.3 & 16.4 & R0010497-081717 & . & 10.2 & 1.94 & 3.19 & 18.7 & 3.57 & --- & --- & 0.2106 & 0.00053 & 4 & 113 & --- & --- & --- & broad\\
g0011247-112844 & 00 11 24.74 & -11 28 43.7 & 14.5 & 14.3 & R0011246-112843 & . & 64.6 & 5.36 & 2.74 & 112 & 9.31 & --- & --- & --- & --- & 0 & 0 & --- & --- & --- & ---\\
g0011304-444115 & 00 11 30.36 & -44 41 15.1 & 17.6 & 17.4 & R0011311-444129 & . & 12.5 & 3.03 & 2.74 & 21.7 & 5.25 & --- & --- & 0.1984 & 0.00020 & 4 & 113 & --- & --- & --- & broad\\
g0011539-323755 & 00 11 53.88 & -32 37 54.8 & 18.6 & 17.7 & R0011527-323808 & . & 7.27 & 1.72 & 1.46 & 10 & 2.37 & --- & --- & 0.2608 & 0.00017 & 4 & 113 & --- & --- & --- & narrow\\
g0012167-054339 & 00 12 16.68 & -05 43 38.8 & 17.5 & 16.5 & R0012163-054335 & . & 5.94 & 1.51 & 3.09 & 10.8 & 2.75 & --- & --- & -9.999 & --- & 2 & 113 & --- & --- & --- & ---\\
\hline    
\end{tabular}
\end{minipage}
  \end{center}
}
\end{table}
\end{landscape}

\noindent the 6dFGS position were rejected and the remainder were inspected by eye. The reader is referred to \citet{mauch} for a detailed description of the radio-optical cross-matching technique. 

\citet{mauch} estimated a reliability of better than 99$\%$ and completeness of better than 96$\%$ for NVSS identifications of 6dFGS galaxies. In this context we use the term `reliability' to refer to the number of spurious identifications obtained (less than 1$\%$) and `completeness' to refer to the number of geniune associations missed (less than 4$\%$). It is expected that SUMSS identifications will have similar reliability and completeness as its resolution at southern declinations is comparable to that of the NVSS. At $\delta>-40^\circ$ where the major axis of the SUMSS beam is $>70$\,arcsec, NVSS and SUMSS contours were simultaneously overlaid onto optical images and inspected to improve the reliability of the cross-matching. Using this method found that 918 ($27\%$) RASS--6dFGS sources are detected in the radio. We define this sample as the RASS--6dFGS `radio sample' and it is discussed in more detail in Section \ref{radioprops}. 

\section{The Data Catalogue}

The final data catalogue contains 3405 RASS sources with corresponding optical and redshift information where available. These are catalogued by their 6dFGS targetname of the form `gHHMMSSS$-$DDMMSS', reflecting the J2000 coordinates. The content of each column are as follows:\\
(1) 6dFGS targetname; of the form gHHMMSSS$-$DDMMSS.
\\
(2--3) Optical position in J2000 coordinates.
\\
(4--5) $b_{\rm J}$ and R magnitudes from the USNO database.
\\
(6) RASS name; of the form RHHMMSSS$-$DDMMSS, reflecting J2000 coordinates.
\\
(7) Extended flag: `E' means that the X-ray source is extended. This is determined such that the source extent given in the RASS--BSC is larger than 35$''$.
\\
(8--9) RASS counts per second and uncertainty. 
\\
(10) Galactic hydrogen column density in units of cm$^{-2}$. This was calculated using the N(H) program in the HEASARC collection which uses data from \citet{1990ARA&A..28..215D} and \citet{2005A&A...440..775K}.
\\
(11--12) RASS flux and uncertainty in erg\,cm$^{-2}$\,s$^{-1}$. The flux was calculated using a fixed photon index of 1.7 and the Galactic column to determine the unabsorbed flux in the 0.1-2.4\,keV band. This made use of the PIMMS program in the HEASARC collection. 
\\
(13) NVSS flux in mJy.
\\
(14) SUMSS flux in mJy.
\\
(15--16) Redshift and uncertainty from the 6dFGS.
\\
(17) Quality flag; `0' signifies that the object wasn't observed as part of the 6dFGS. Values $\geq 3$ are regarded as reliable; See Section \ref{6dfspectra}.
\\
(18) Programme ID (progID) of the 6dFGS spectrum. See Section \ref{progid}. A value of `0' once again signifies that the source wasn't observed.
\\
(19) Classification from NED
\\
(20) Redshift from NED
\\
(21) Reference of the NED redshift.
\\
(22) Comments - This is a brief note regarding any features in the 6dFGS spectrum. See Section \ref{comments}.
\\
We have chosen to include the USNO magnitudes in the catalogue as these were originally used to select targets prior to the 6dFGS observations. Optical magnitudes from SuperCOSMOS have since been included as part of the 6dFGS target sample and are available from the 6dFGS database \citep{6dfDR2, 6df2009}. The first 50 entries of the RASS--6dFGS catalogue are shown in Table \ref{examplecat}.  

\subsection{Comments in the 6dFGS-RASS catalogue} \label{comments}

To prevent these comments from being unneccesarily complicated they are generally self-explanatory, one-word entries as follows: \\ 
`broad' - the spectrum features broad emission lines. \\
`narrow' - the spectrum has narrow emission features. \\
`abs' - the spectrum exhibits only absorption lines. \\
`BLLac' - a featureless spectrum with strong continuum.  \\
`active M-star' - characteristic M-star spectrum with strong Balmer emission at z=0. \\
`WD' - white dwarf star.\\
`CV' - Cataclysmic variable star.\\
`neb' - Spectrum displays $z=0$ nebula emission lines.\\
`bad splicing' - refers to an error in the data reduction process when the two arms of the spectrum (blue and red) were not matched together correctly in the final spectrum.\\
`fringing' - fringing occasionally occurs causing strong oscillations in the spectrum. This is due to either a small air pocket or damage in the fibre which causes it to act like a fabry-perot filter. Some redshifts could still be distinguished accurately.\\
`blue/red arm only' - an error with either the red or blue arm during observations resulting in only half of the spectrum being available. \\
`contamination' - a nearby source (generally a foreground star) dominates the spectrum, masking any optical signature of the RASS source.\\ 
A question mark signifies any uncertainty in the comments. These comments were added during the visual inspection of the 6dFGS redshifts and should be used as a guide only. 

\section{Discussion}

\subsection{Redshift Distributions}

\begin{figure}
  \centerline{\epsfig{file=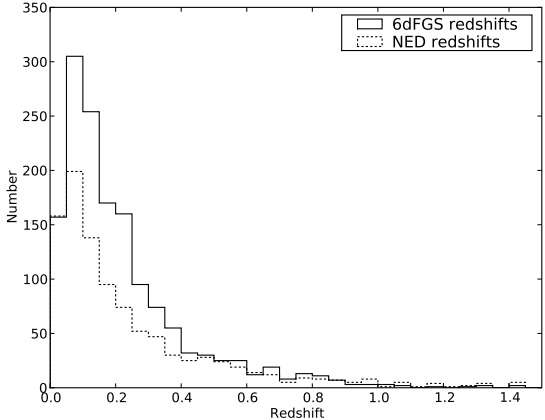, width=\linewidth}}
  \caption {RASS--6dFGS redshift distribution, excluding Galactic sources. The solid line is the redshift distribution from the 6dFGS spectroscopic sample and the dotted line is the distribution of redshifts found in the literature. There are an additional 9 6dFGS and 15 NED sources with $z\geq1.5$ not shown. }
\label{zhist}
\end{figure}

The distribution of redshifts in the RASS--6dFGS catalogue is displayed in Figure \ref{zhist} for all the extragalactic sources. The median redshift for the extragalactic RASS--6dFGS sample is $z=0.1557$ and the mean redshift is $\overline{z}=0.2264$. The median redshift from the literature is fairly similar at $z=0.1550$ while the mean is slightly higher at $\overline{z}=0.2718$. A clear peak is evident in both distributions at around $z=0.1$. These results are consistent with the median and mean redshifts of $z=0.1218$ and $\overline{z}=0.2100$ found in the extragalactic RBSC--NVSS sample \citep{bauer}. A more detailed comparison of these catalogues is discussed in Section \ref{rbsc-nvss}.

\begin{figure}
  \centerline{\epsfig{file=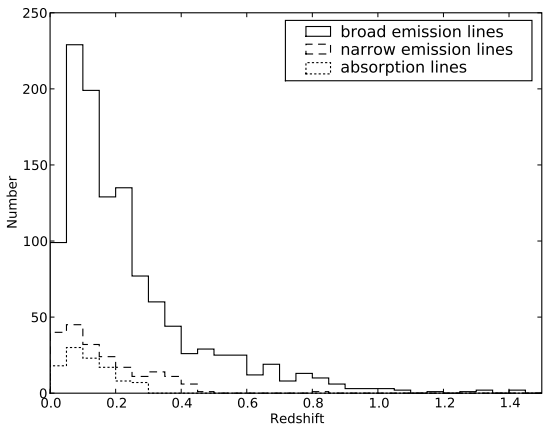, width=\linewidth}}
  \caption {Redshift distribution of the RASS--6dFGS spectroscopic sample, divided into 3 main classes according to their spectral features. Again there are an additional 9 sources with $z\geq1.5$ not shown (these sources all exhibit broad emission lines). }
\label{zlineshist}
\end{figure}

Of the sources in the spectroscopic sample, 1171 show broad emission features and 202 display narrow emission lines implying that approximately 68$\%$ are Type 1 AGN and 12$\%$ are Type 2. The remaining sources are generally either absorption-line galaxies (6$\%$) or stars (13$\%$). Figure \ref{zlineshist} shows the redshift distributions of each of these classes (excluding the stars); sources with broad emission lines are shown by the solid line, objects with narrow emission lines by the dashed line and absorption line galaxies by the dotted line. The median redshifts for each of these classes are $z=0.1706$, $z=0.1291$ and $z=0.1138$ respectively.

\subsection{Optical Magnitude Distribution}

\begin{figure}
  \centerline{\epsfig{file=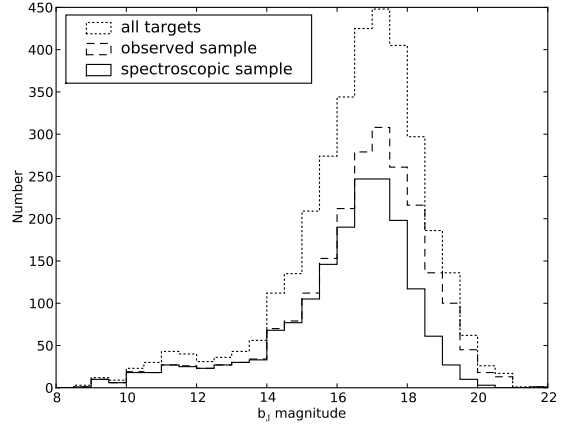, width=\linewidth}}
  \caption{$b_{\rm J}$ magnitude distribution. Optical magnitudes were taken from the USNO--A2.0 catalogue.} 
 \label{bmaghist}
\end{figure}

Figure \ref{bmaghist} shows the distribution of $b_{\rm J}$ magnitude for the RASS--6dFGS catalogue. The dotted line shows the distribution of the full catalogue, while the dashed and solid lines show the magnitude distribution of the observed sample, and the spectroscopic sample respectively. This distribution peaks at around $b_{\rm J}$ $\sim17$ with a median magnitude of $b_{\rm J}$=16.8. Kolmogorov-Smirnov tests show that we can reject the null hypothesis that the spectroscopic sample is representative of both the full sample and the observed sample at a significance level of 99.9$\%$. On the other hand, K-S tests show that the RASS--6dFGS full catalogue and the observed sample are mutually compatible and representative of the same population. This is in agreement with Figure \ref{completenessbmagfig} which, due to the flat distribution of the observed sample, shows that the selection of sources observed is not dependant on optical magnitude. 

Looking at just the spectroscopic sample reveals a median $b_{\rm J}$ magnitude of 16.6. When split according to their spectral line properties the median optical magnitudes remain fairly similar with $b_{\rm J}=16.6$, 16.7 and 16.3 for the broad emission, narrow emission and absorption line objects respectively. 

\begin{figure*}
  \centerline{\epsfig{file=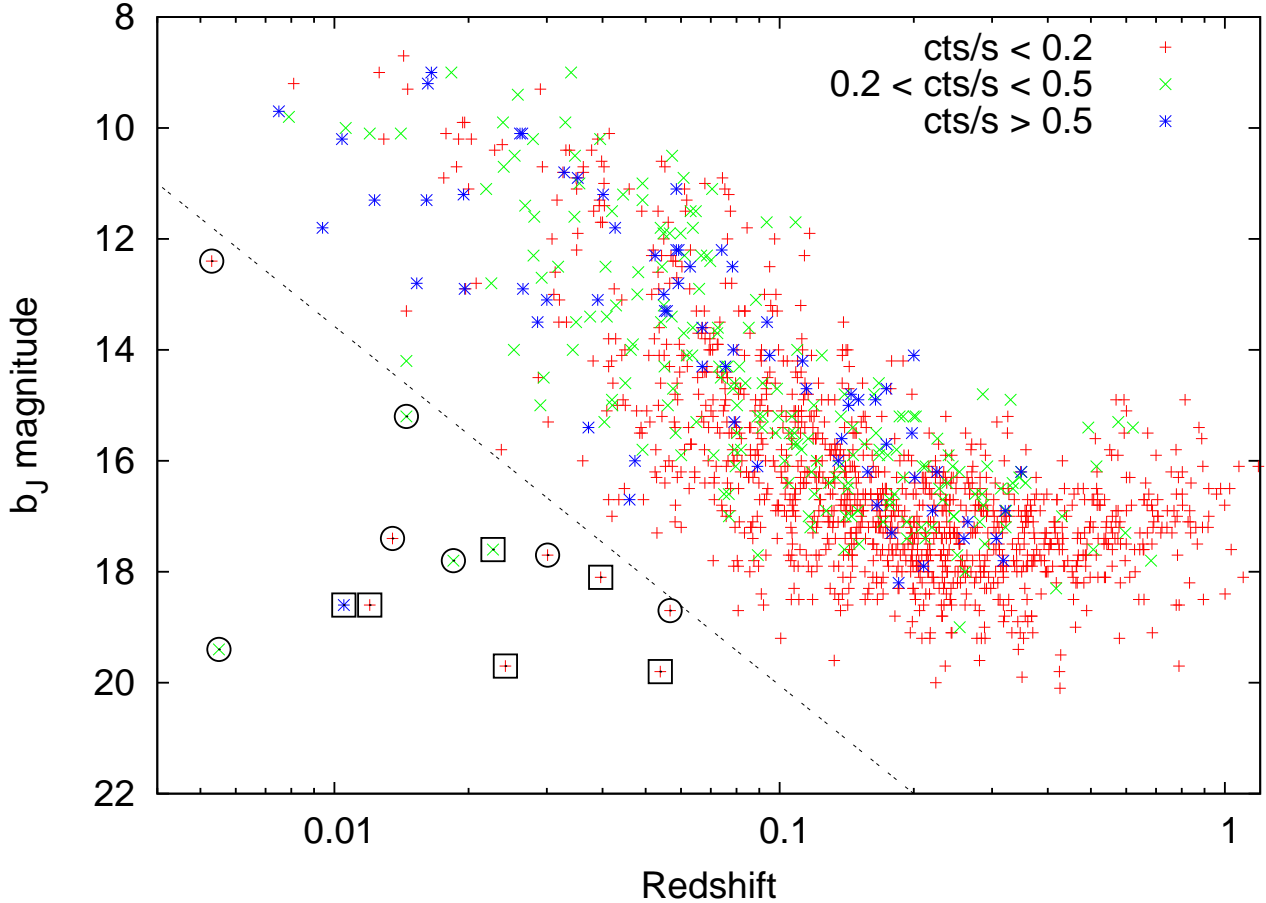, width=\linewidth}}
  \caption{$b_{\rm J}$ magnitude against redshift for sources with reliable 6dFGS redshifts. Red `+' marks sources with a RASS count rate $< 0.2$\,cts/s, green `x' are sources with 0.2 $\leq$ count rate $< 0.5$\,cts/s and the blue `*' are sources with a count rate $\geq 0.5$\,cts/s. Sources that lie below the dashed line are outliers and are discussed in Section \ref{outliers} and Table \ref{outlierstab}. The circles mark objects that fall into case (a) and the squares mark objects that fall into case (b). This line was set at a point such that only the most extreme outliers were selected.}
 \label{bmagvsz}
\end{figure*}

Figure \ref{bmagvsz} plots $b_{\rm J}$ magnitude against redshift for objects in the spectroscopic sample, excluding the Galactic sources. These sources were also divided into three groups according to X-ray count rates. The red `+' corresponds to sources with a count rate $< 0.2$\,cts/s, green `x' marks the sources with a count rate between 0.2 and 0.5\,cts/s and the blue `*' are sources with a count rate above 0.5\,cts/s. 

As expected there is a correlation between optical magnitude and redshift at lower redshifts where the galaxies are the dominant population. Towards higher redshifts where the QSOs begin to dominate this correlation begins to flatten out. Sources that appear not to follow this trend are discussed in Section \ref{outliers}. The brighter X-ray sources are somewhat concentrated towards the region of optically bright and nearby galaxies, while the fainter X-ray sources are generally more distant and optically fainter. 

\subsection{Outliers in the magnitude-redshift distribution} \label{outliers}

Figure \ref{bmagvsz} revealed 13 sources which do not follow the trend of becoming optically fainter with distance. Although there is a broad distribution of luminosities, particularly at low redshift, these objects fall significantly outside the range expected (marked by the dashed line). Investigating the 6dFGS spectra of these sources revealed that these sources fall into two main categories: (a) The object is a large, very nearby galaxy making it difficult to accurately calculate the optical magnitude. These are circled in Figure \ref{bmagvsz}. The USNO optical magnitude is most likely underestimated for these sources. (b) The object appears compact in the DSS images and the spectra is on average noisier than the majority of 6dFGS spectra. These sources are marked by a square in Figure \ref{bmagvsz}. Table \ref{outlierstab} lists the reason for each source.

\begin{table*}
\begin{center}
\caption{RASS--6dFGS sources that are outliers in Figure \ref{bmagvsz}. B$_{\rm J}$ magnitudes from the SuperCOSMOS database have been included for comparison and absolute magnitudes have been derived from these magnitudes. The SuperCOSMOS class refers to the optical morphology of the source on the UKST plate; 1 = extended, 2 = stellar and 4 = noise. } 
\label{outlierstab}
\begin{tabular}{lccccccl}
\hline
\bf{6dFGS name} & \bf{USNO} & \multicolumn{3}{c}{\bf{SuperCOSMOS}} & \multicolumn{2}{c}{\bf{6dFGS}} & \bf{Reason for discrepancy}\\
& $b_{\rm J}$ & B$_{\rm J}$ & M$_{\rm B}$ & Class & z & q &\\
\hline
\multicolumn{8}{l}{(a) These are marked by circles in Figure \ref{bmagvsz}: }\\
g0129067-073830 & 18.7 & 16.0 & -20.9 & 1 & 0.0567 & 4 & Nearby galaxy. SuperCOSMOS magnitude is more reliable.\\
g0153005-134419 & 12.4 & 12.5 & -19.1 & 1 & 0.0053 & 4 & Nearby elliptical galaxy. \\
g0342037-211440 & 15.2 & 7.2 & -26.6 & 1 & 0.0145 & 4 & Nearby galaxy. SuperCOSMOS magnitude is more reliable. \\
g0650175-380514 & 17.7 & 16.1 & -19.4 & 1 & 0.0301 & 4 & Nearby Seyfert galaxy. \\
g1256101-080905 & 17.4 & 9.1 & -24.6 & 2 & 0.0135 & 4 & Nearby face-on spiral. SuperCOSMOS magnitude is more reliable.\\
g1942406-101925 & 19.4 & 7.2 & -24.5 & 2 & 0.0055 & 4 & Nearby Seyfert galaxy. SuperCOSMOS magnitude is more reliable.\\
g2359107-040737 & 17.8 & 14.0 & -20.4 & 4 & 0.0185 & 4 & Nearby Seyfert galaxy. \\
\multicolumn{8}{l}{(b) These are marked by squares in Figure \ref{bmagvsz}: }\\
g0014114-502235 & 18.6 & 19.4 & -13.7 & 2 & 0.0105 & 3 & Unknown; see Section \ref{outliers}. \\
g0200209-410936 & 19.8 & 20.3 & -16.5 & 2 & 0.0539 & 4 & Unknown; see Section \ref{outliers}.  \\
g0245228-240617 & 19.7 & 20.3 & -14.7 & 2 & 0.0242 & 3 & Unknown; see Section \ref{outliers}.  \\
g0410498-272959 & 18.6 & 19.2 & -14.2 & 2 & 0.0120 & 4 & Possible extragalactic HII region. \\
g1513242-075452 & 18.1 & 18.6 & -17.5 & 1 & 0.0396 & 3 & Uncertain redshift. \\
g2123073-103649 & 17.6 & 18.0 & -16.8 & 1 & 0.0227 & 4 & Unknown; see Section \ref{outliers}.  \\
\hline
\end{tabular}
\end{center}
\end{table*}

\begin{figure*}
  \centerline{\epsfig{file=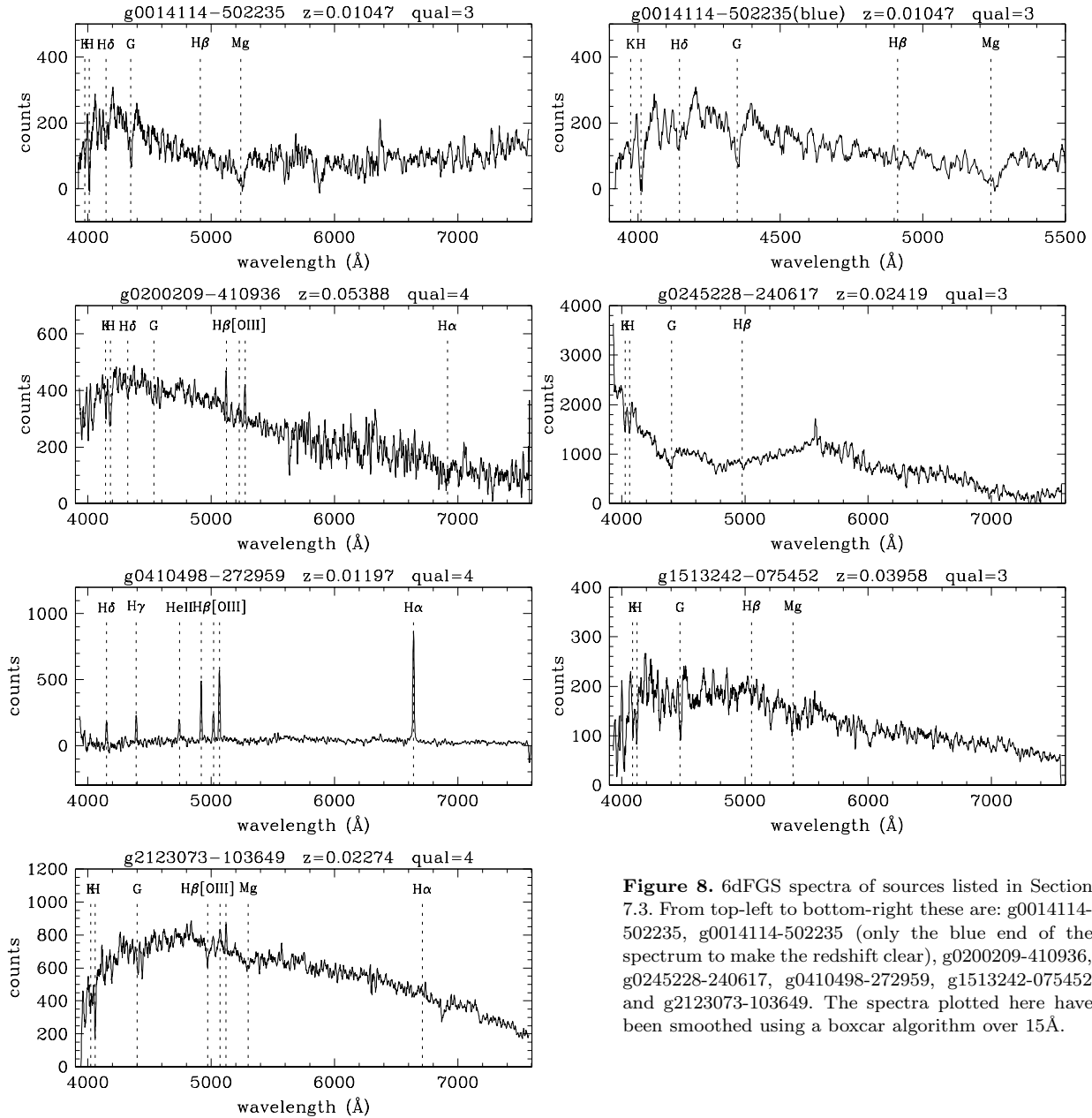, width=\linewidth}}
\hspace{9.5cm}
\begin{minipage}{0.4\linewidth}
\vspace{-6cm}
   \caption{6dFGS spectra of sources listed in Section \ref{outliers}. From top-left to bottom-right these are: g0014114-502235, g0014114-502235 (only the blue end of the spectrum to make the redshift clear), g0200209-410936, g0245228-240617, g0410498-272959, g1513242-075452 and g2123073-103649. The spectra plotted here have been smoothed using a boxcar algorithm over $15\rm{\AA}$. \label{outliersspectra}}
\end{minipage}
\end{figure*}

For the sources that appear to have correct optical magnitudes (case (b)) 6dFGS spectra are shown in Figure \ref{outliersspectra} and in most cases the redshifts also appear to be correct. These objects are listed below along with information available from existing databases. Further observational follow-up is needed to confirm the nature of these objects. Approximate X-ray luminosities have been calculated using the 6dFGS redshifts and assuming a cosmology with H$_0=75$\,km\,s$^{-1}$\,Mpc$^{-1}$, $\Omega_M=0.3$ and $\Omega_\Lambda=0.7$. \\
\noindent{\it g0014114-502235:} Although the spectrum is reasonably noisy, CaII K$\&$H and G bands are evident suggesting the redshift is reliable.  Searching NED revealed that this source is listed as a BL-Lac in \citet{VCV}, but no redshift is given. There is a GALEX \citep{2005ApJ...619L...1M} detection with FUV$=19.9$\,mag and NUV$=19.4$\,mag as well as a SUMSS flux of 15.1\,mJy. We calculate an X-ray luminosity of $4\times 10^{42}$\,erg/s which is typical of the luminosity expected in an AGN. However, at this redshift it is very unusual that the host galaxy appears point-like in the optical images. \\
{\it g0200209-410936:} Once again this redshift is reliable with CaII K$\&$H absorption lines, H$\beta$ and [OIII] emission lines. It is strange that there is no H$\alpha$ emission which would be expected due to the presence of H$\beta$ emission, but there is an atmospheric absorption band at $\sim 6900 \rm{\AA}$ which could account for this. This redshift gives an approximate X-ray luminosity of $1\times 10^{43}$\,erg/s which is too luminous to be a dwarf or starburst galaxy, HII region or even an ULX. There is a GALEX detection (FUV=20.2, NUV=20.0\,mag) and a faint radio source (5.9\,mJy) determined from the SUMSS images. \\ 
{\it g0245228-240617:} There is a small error in the splicing of the optical spectrum as the continuum appears to peak at approximately 5500\,$\rm{\AA}$. Ignoring this there are still many counts in the continuum, particularly in the blue end of the spectrum, but no emission or absorption features suggesting this could be a BL-Lac object. There is a faint GALEX detection with FUV=21.1 and NUV=20.2\,mag. At this redshift the X-ray luminosity is approximately $2\times 10^{42}$\,erg/s. \\
{\it g0410498-272959:} The 6dFGS spectrum displays strong emission features typical of an extragalactic HII region. However, in addition there is a strong HeII4686 emission line rarely seen in HII regions. The high energy photons required for this emission (the ionization potential of HeII is 54\,eV) argues that the X-ray counterpart is correct. The spectrum also does not display any [NII] emission suggesting that this source has very low metallicity. There is a GALEX source (FUV=20.5, NUV=20.4\,mag) but no known radio detection. The X-ray luminosity is $4\times 10^{41}$\,erg/s making it a very luminous extragalatic HII region. \\
{\it g1513242-075452:} This source has an NVSS flux of 10.7\,mJy. The optical image shows two sources close together on the sky (separated by $\sim$3\,arcsec) making it difficult to determine the correct counterpart. In fact, since the diameter of the 6dF fibres are approximately 6\,arcsec we also cannot determine if the optical spectrum is of one particular source or a combination of both. The first step would be to confirm the redshift. \\
{\it g2123073-103649:} This object also appears in the RBSC--NVSS sample with the same redshift ($z=0.0227$) and NVSS flux of 118.8\,mJy. It has a GALEX detection with FUV=21.38 and NUV=18.41 and an X-ray luminosity of $\sim 1\times 10^{43}$\,erg/s which is indicative of an AGN. \\

\subsection{Radio properties of RASS--6dFGS sources} \label{radioprops}

There are 918 (27$\%$) sources that have radio counterparts in either the NVSS or SUMSS catalogues as described in Section \ref{radiodetections}. The NVSS detects 710 RASS--6dFGS sources (30$\%$) above its flux limit of $S\sim2.5$\,mJy north of $\delta=-40^\circ$. SUMSS detects 309 RASS--6dFGS sources (19$\%$) above $S=6$\,mJy south of $\delta=-30^\circ$. There are 101 sources in the region between $\delta=-30^\circ$ and $\delta=-40^\circ$ where the 843\,MHz SUMSS and 1.4\,GHz NVSS surveys overlap. These sources have a median radio spectral index of $\alpha=-0.49$ ($S\propto\nu^\alpha$) which is flatter than the median spectral index of $\alpha=-0.83$ measured from all SUMSS and NVSS sources in this region \citep{sumss}. The flatter median radio spectral index of RASS--6dFGS sources is probably caused by the larger fraction of QSOs detected by the X-ray selected sample, whereas the full SUMSS and NVSS catalogues are dominated by radio galaxies with steeper spectral indices.

For the spectroscopic sample the radio detection rate is 24$\%$, making it statisically consistent with the detection rate of the full sample. Dividing these according to their spectral line properties reveals that 24$\%$ of the broad emission line sources have radio counterparts, 33$\%$ for the narrow emission line sources and 59$\%$ for the absorption line objects.

\begin{figure}
  \centerline{\epsfig{file=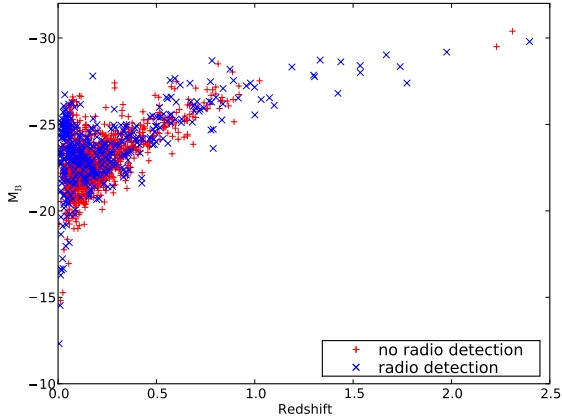, width=\linewidth}}
  \caption{Absolute magnitude as a function of redshift for the subset of RASS--6dFGS sources in the spectroscopic sample.} 
  \label{radioabsmag}
\end{figure}

\begin{figure}
  \centerline{\epsfig{file=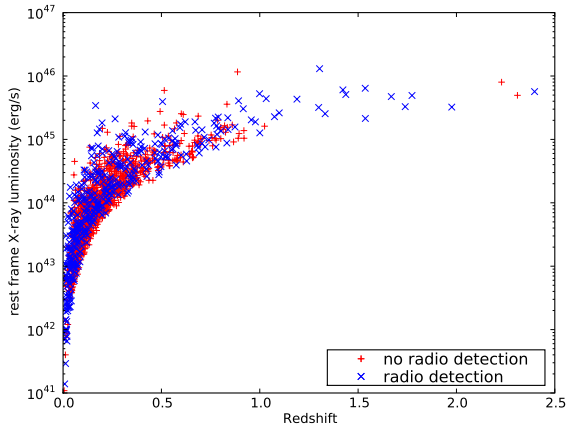, width=\linewidth}}
  \caption{K-corrected X-ray luminosity as a function of redshift. To calculate these luminosities we have used a spectral index of -0.7. } 
  \label{radioxray}
\end{figure}

Figures \ref{radioabsmag} and \ref{radioxray} show both absolute magnitude and X-ray luminosity as a function of redshift for both the sources with radio detections and those without. What is immediately obvious in these plots is that at high redshifts virtually all of the sources fall into the radio sample and in fact are quite strong radio sources\footnote{Searching the literature revealed that the two sources at high redshift without radio detections (g1106335-182124 and g2026104-453627) are gravitational lenses and hence possibly only appear in our catalogue due to the X-ray and optical emission being magnified.}. The median flux density for sources with $z>1$ is 1151\,mJy while the median flux density for the entire RASS--6dFGS radio sample is 28.6\,mJy. This suggests that the more distant sources are Doppler boosted due to the presence of a radio jet pointed towards our line of sight. As such, the X-ray emission (or some fraction of it) is also being boosted by a jet component and hence we only detect bright X-ray sources that are radio-loud at high redshifts. This is in agreement with \citet{1987ApJ...313..596W} who found that radio-loud QSOs have a higher average X-ray luminosity.

\begin{figure}
  \centerline{\epsfig{file=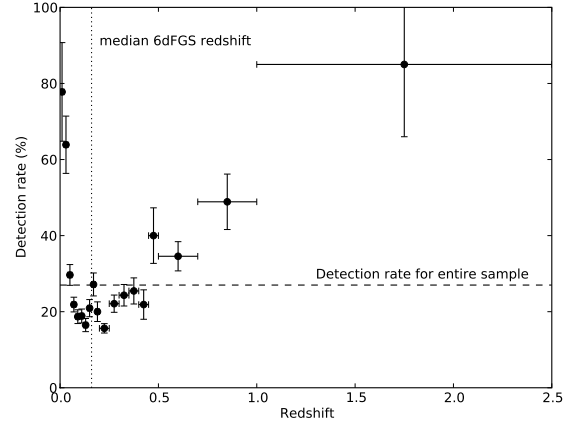, width=\linewidth}}
  \caption{The radio detection rate against redshift. The vertical errorbars represent $\sqrt n$ counting errors while the horizontal errobars represent the size each of the bins. } 
  \label{radiodetz}
\end{figure}

The fraction of sources with radio detections changes with redshift, as shown in Figure \ref{radiodetz}. There is a high detection rate at low redshift where we detect a large fraction of low luminosity, radio-quiet AGN in addition to the radio-loud sources. This drops off quite rapidly as these radio-quiet sources fall below the detection limit of the radio surveys, leaving only the radio-loud AGN in the sample. The detection rate then begins to rise again towards higher redshifts as the number of radio-loud sources increases. 

In contrast to our X-ray selected sample, high redshift optical quasar samples, even at the brightest magnitudes, find that only $\sim20-25$ percent of the objects are radio loud (e.g. \citet{2007ApJ...656..680J}). This may go some way to explaining the flatter bright end slope of the QSO luminosity function derived from the X-ray (e.g. \citet{2005A&A...441..417H}), as compared to the optical (\citet{2006AJ....131.2766R}; \citet{croom09}). The number of bright X-ray quasars is boosted by a population which is dominated by X-ray emission from the jet, presumably beamed at some level. This is also consistent with both \citet{1998MNRAS.300..625W} and \citet{2005MNRAS.357.1267C} who find a flatter slope in the radio luminosity function for a sample of radio-loud QSOs.

Since the RASS--6dFGS catalogue was selected to be the brightest X-ray sources over a large area of sky we would only expect to see this effect in a sample such as this one. Deeper X-ray surveys would not observe this due to the small area of sky covered and the rarity of these extremely luminous sources. As such, the bright end of the luminosity function is not constrained by these surveys. 

K-S tests comparing the redshift and optical magnitude distributions of the full sample and the radio sample show that they are different populations at a significance level of 99$\%$. This highlights the need to study the full RASS--6dFGS sample as opposed to just the sources with radio counterparts. 

\subsection{RASS Targets in other 6dFGS Additional Samples} \label{progid}

The 6dF Galaxy Survey takes its targets from a number of different samples. These were all given a number referred to as the programme ID (progID) and then listed in order of priority \citep{6df}. Sources selected from the {\it ROSAT} All Sky Survey were designated a progID of 113. However, many of the targets were also listed in other 6dFGS target samples. Due to this the progID of each spectrum refers to the target sample of the highest priority, where the higher priorities carried greater importance in the allocation of fields; i.e. if a target is also listed in the primary sample then it is assigned a progID number of 1 since that target list had the highest priority. Table \ref{progidtab} shows the different progID numbers listed in the RASS--6dFGS catalogue. The majority first appear in the RASS target sample\footnote{These sources could also appear in samples with lower priorities.}, with 15.4$\%$ of RASS targets also appearing in the primary sample, selected from the 2MASS eXtended Source Catalogue (2MASS XSC).  

\begin{table}
\begin{center}
\caption{List of 6dFGS target samples, progID and number of sources. The second entry is the primary sample, followed by the additional samples. Again the percentages are given with respect to the RASS--6dFGS full catalogue and to the observed sample.}
\label{progidtab}
\begin{tabular}{lcccc}
\hline
\bf Survey	& \bf Prog & \bf No. & \bf $\%$ & \bf $\%$\\	
& \bf ID &  & Full & obs. \\ 
\hline
Not Observed & 0 & 1181 & 34.7 & --- \\
2MASS K$<$ 12.75 & 1 & 341 & 10.0 & 15.4\\
SuperCOSMOS $r_{\rm F}<$ 15.6 & 7	& 4 & 0.1 & 0.2 \\ 
SuperCOSMOS $b_{\rm J}<$ 16.75 & 8	& 2 & 0.1 & 0.1 \\
IRAS FSC ($6 \sigma$) & 126	& 17 & 0.5 & 0.7\\
ROSAT All-Sky Survey & 113	& 1860 & 54.6 & 83.6\\
\hline
Total &  &  3405 & & \\
\hline
\end{tabular}
\end{center}
\end{table}

\subsection{Comparison with the RBSC--NVSS catalogue} \label{rbsc-nvss}

In order to compare our RASS--6dFGS catalogue with the RBSC--NVSS sample \citep{bauer}, a subset of objects were selected from each in the overlapping region of sky. This involved selecting only the sources in the declination range $-40^{\circ}<\delta<0^{\circ}$, $|b|>15^{\circ}$ and counts/s$>0.1$. This leaves 609 sources in the RBSC--NVSS comparison catalogue and 956 in the RASS--6dFGS comparison catalogue. However, only 387 sources appear in both. The 222 sources that appear in the RBSC--NVSS comparison sample but not in the RASS--6dFGS catalogue are the candidate galaxy clusters which were removed in our selection process. 94 of these sources were observed as part of the entire 6dF Galaxy Survey and as a result optical spectra of these are available from the 6dFGS website, but are not included in this catalogue. These sources are discussed in Section \ref{clusters}. 

There are 569 (59.5$\%$) sources that appear in the RASS--6dFGS comparison catalogue, but were not found in the RBSC--NVSS sample. The main reason for this (accounting for 96$\%$) is that these objects do not have radio counterparts above $S_{1.4}=2.5$\,mJy, the NVSS flux limit. 

\begin{figure}
  \centerline{\epsfig{file=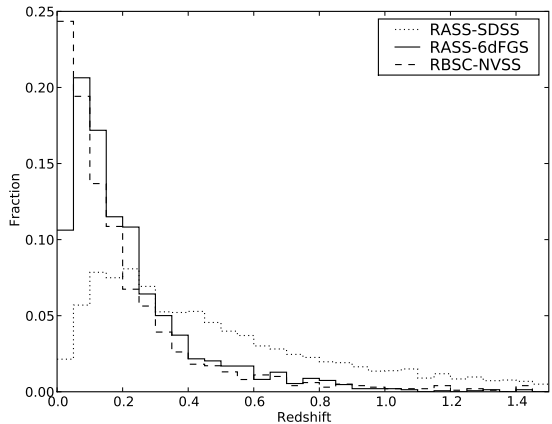, width=\linewidth}}
  \caption{Redshift distribution of RASS--6dFGS sources (solid line) compared to the RBSC--NVSS sample (dashed line) and the RASS--SDSS AGN catalogue (dotted line). Each distribution has been normalised by the total number of sources in the catalogue. Once again sources with $z\geq 1.5$ are not shown.} 
 \label{allzhist}
\end{figure}

\begin{figure}
  \centerline{\epsfig{file=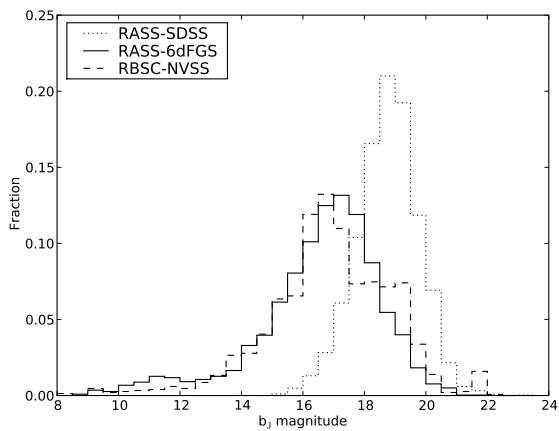, width=\linewidth}}
  \caption{$b_{\rm J}$ magnitude distribution of RASS--6dFGS sources (solid line) compared to the RBSC--NVSS sample (dashed line) and the g magnitude distribution of the RASS--SDSS AGN catalogue (dotted line). Again these have been normalised by the number of sources in each catalogue in order to compare them.} 
 \label{allbmaghist}
\end{figure}

Figures \ref{allzhist} and \ref{allbmaghist} show that although the RASS--6dFGS and RBSC--NVSS samples were selected in slightly different ways, the redshift and magnitude distributions are reasonably similar, with the only significant difference being the number of objects in the RASS--6dFGS sample. 


\begin{figure}
  \centerline{\epsfig{file=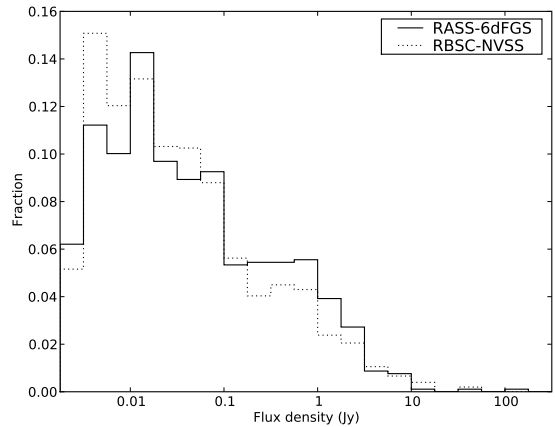, width=\linewidth}}
  \caption{Distribution of 1\,GHz flux density for the RASS--6dFGS sample (solid line) and the RBSC--NVSS sample (dotted line). Here the flux density has been binned in log flux and normalised by the number of sources in each catalogue.} 
 \label{radiodist}
\end{figure}

As only $27 \%$ of our sample are detected in the radio, it highlights how many potential AGN are missed in a radio selected sample. The RASS--6dFGS radio sample (when tailored to the selection criteria described above) can be viewed as the southern counterpart to the RBSC--NVSS catalogue. Figure \ref{radiodist} shows that the distribution of flux density at $\sim$1\,GHz is very similar for these two samples. In this figure, the observed flux densities in each catalogue were firstly binned in log flux and then normalised by the total number of sources of each sample. Looking at the fraction of sources in each bin allows us to compare these samples more accurately. The slight excess of RBSC--NVSS sources in the lower bins is due to the marginally fainter flux limit of the NVSS (2.5\,mJy) as opposed to the SUMSS catalogue (6\,mJy).

Kolmogorov-Smirnov tests comparing the optical magnitude and redshift distributions of the RBSC--NVSS sample and the RASS--6dFGS radio sample show, with 99$\%$ significance, that we can reject the null hypothesis that they are drawn from the same population. However, reducing the RASS--6dFGS sample to only those sources with an X-ray count rate above 0.1 cts/s and repeating the K-S tests reveals that we can no longer reject this null hypothesis (the probability of rejection is 83.3$\%$). The optical magnitude distributions remain different at the 99$\%$ significance level due to slightly larger fraction of optically bright objects in the RASS--6dFGS catalogue.

\subsection{6dFGS spectra of RASS cluster galaxies} \label{clusters}

A number of RASS sources were removed from the 6dFGS target list under the assumption that they were cluster galaxies. This makes it unclear if the X-ray emission originates from an AGN or from the intracluster medium. Comparisons with the RBSC--NVSS catalogue revealed 222 sources that were excluded due to this selection criteria, but a number were observed as part of the 6dF Galaxy Survey, the majority as part of the primary 2MASS selected survey. Of the 94 sources with optical spectra, 81 have reliable redshifts. These were visually inspected to check if a large number of AGN were excluded in the selection process. Table \ref{clusterspectratab} details this inspection process. As expected, the majority (78$\%$) are absorption line objects, typical of a passive galaxy, with the optical spectrum showing no features of AGN activity. A small number of sources display emission lines, but 4 of these are starforming galaxies based on the [NII]$\lambda$6583/H$\alpha$ and [OIII]$\lambda$5007/H$\beta$ ratios \citep{2001ApJ...556..121K}. 

Excluding these, along with all the absorption line objects, leaves only 12 (14.8$\%$) sources where an AGN is indicated by the optical spectrum. However, the low spatial resolution of RASS images makes it difficult to determine what fraction of X-ray emission originates from the AGN and what fraction may be attributed to the intracluster medium. This can be compared to both the full RASS--6dFGS catalogue, and the subset of sources which were also observed as part of the primary survey (2MASS selected targets) as shown in Table \ref{clusterspectratab}. While less than 15$\%$ of the cluster galaxies show optical signatures of an AGN, $\sim 80\%$ of the RASS--6dFGS sources and $\sim 90\%$ of those observed as part of the primary survey (progID=1) display emission lines indicative of AGN activity. 

This confirms that removing the candidate cluster galaxies in the selection process does not eliminate a large number of AGN. By removing these cluster galaxies it also means that we can assume that the majority of the X-ray emission is due to an AGN for our catalogued objects. 

\begin{table}
\begin{center}
\caption{Results from the visual inspection of optical spectra of the cluster galaxies excluded in the selection process. Only a small fraction of these are AGN. This contrasts greatly to both the RASS selected targets (column 4) and the targets that were also part of the primary sample (column 5).} \label{clusterspectratab}
\begin{tabular}{lcccc}
\hline
\bf{Features evident} & \multicolumn{2}{c}{\bf Clusters} & \bf Full & {\bf progid$=1$} \\
 & \bf{No.} & \bf{$\%$} & \bf{$\%$} & \bf{$\%$} \\
\hline
Absorption lines & 64 & 79.0 & 6.4 & 11.5 \\
Narrow emission lines & 9 & 11.1 & 11.8 & 17.4 \\
Broad emission lines & 7 & 8.7 & 67.9 & 71.1 \\
Star & 1* & 1.2 & 13.9 & 0\\
\hline
\end{tabular}
\end{center}
*{\scriptsize This classification is due to an optical misidentification or a foreground star dominating the spectrum.}
\end{table}

\subsection{RASS selected AGN with SDSS spectroscopy}

\citet{2003AJ....126.2209A, 2007AJ....133..313A} present a catalogue of X-ray selected AGN from the {\it ROSAT} Bright and Faint Source Catalogues with optical spectra in the SDSS Data Release 5. This catalogue currently covers a magnitude range of $15<g<23$ and a redshift range of $0.01<z<4$, with median values of $g=18.8$ and $z=0.42$ respectively. The majority of RASS identifications are quasars and Seyfert 1 galaxies, but there are also a small number of BL-Lac candidates. Of the 503 objects from this catalogue that are in the southern sky, 126 (25$\%$) are also in our RASS--6dFGS sample. The RASS Faint Source Catalogue \citep{2000IAUC.7432....3V} consists of 105,924 sources, making their combined X-ray sample more than 6 times larger than the Bright Source Catalogue alone, and accounts for the relatively small fraction of overlap. 

Figures \ref{allzhist} and \ref{allbmaghist} clearly show that while the RBSC--NVSS is similar to the RASS--6dFGS catalogue, the SDSS sample is distinctively different. The magnitude distribution\footnote{\citet{2005MNRAS.360..839R} find a median difference of $(g-b_{\rm J})=-0.045$ for $b_{\rm J}$ magnitudes ranging from approximately $b_{\rm J}=18.5$ to $b_{\rm J}=21$. As this difference is so small on the scales we are plotting, no correction has been made here.} in particular is much narrower and peaks at a fainter magnitude of $g\sim19$ while the redshift distribution peaks at a slightly higher redshift ($z\sim0.2$ as opposed to $z\sim0.1$ for the other two catalogues). The RASS--SDSS catalogue contains fainter, more distant sources resulting from the fact that the selection of sources was deeper in both the optical and X-ray regimes. Our RASS--6dFGS catalogue complements this sample by providing uniform optical spectroscopy for the optically brighter X-ray sources in the south.

\section{Conclusions}

In this paper we have presented a catalogue of 3405 X-ray selected sources in the 6dF Galaxy Survey. Sources were selected from the {\it ROSAT} All Sky Survey Bright Source Catalogue, hence are count-rate limited at $0.05$\,cts\,s$^{-1}$. Selection criteria of the 6dFGS means that these sources are also south of $\delta = 0^{\circ}$ and outside the Galactic plane ($|b|>10^{\circ}$). Since the RASS sources were additional targets, not all objects in the catalogue were observed. However, for completeness, X-ray and available optical information have been included in the catalogue for all targets. 

A total of 2224 (65.3$\%$) objects were observed as part of the survey, with the majority (1715 sources) having reliable redshift measurements. For the optically bright objects ($b_{\rm J} \leq 17.5$) in the observed sample, 1333 out of 1478 sources (90.2$\%$) have reliable redshifts. Inspecting the optical spectra of the RASS--6dFGS spectroscopic sample revealed that  1171 sources (68$\%$) exhibit broad emission features, indictative of Type 1 AGN, while only 202 (12$\%$) display narrow emission feaures. The remaining objects are either absorption-line galaxies (6$\%$) or stars (13$\%$). The median redshift of the spectroscopic sample is $z=0.16$, with median redshifts of $z=0.1706$, $z=0.1291$ and $z=0.1138$ for each of the broad emission, narrow emission and absorption line classes respectively.  

We also find a small number of optically-faint, very low redshift, compact objects which fall outside the general trend in the $b_{\rm J}-z$ plane. The X-ray luminosities of these sources range from $4\times 10^{41}$\,erg/s to $1\times 10^{43}$\,erg/s suggesting that they are either very luminous starbursts or AGN. However, in all but one case, the 6dFGS optical spectra does not provide any evidence of AGN or starburst activity. Further observational follow up is needed to confirm the physical properties of these sources. 

There are 918 (27$\%$) RASS--6dFGS sources detected in either the 1.4\,GHz NRAO VLA Sky Survey (NVSS) or the 843\,MHz Sydney University Molonglo Sky Survey (SUMSS). The fraction of sources with radio counterparts changes with redshift, and at $z>1$ nearly all the RASS--6dFGS sources have radio detections. These sources are strong radio sources with a median flux density of 1151\,mJy whereas the median flux density for the full sample is 28.6\,mJy. We attribute this to the presence of a radio jet which Doppler boosts the radio emission. The X-ray flux of these sources is also boosted by a jet component and thus at large redshifts selecting bright X-ray sources preferentially selects radio-loud AGN. 

The RASS--6dFGS catalogue, when reduced to just those sources with radio detections, can be used as the southern counterpart to the RBSC--NVSS catalogue and as such offers a large sample of BL-Lac and blazar sources. Other properties of RASS--6dFGS sources, in particular the high-frequency radio properties, will be examined in forthcoming papers. This will enable further studies of BL-Lac objects and a more extensive analysis of the multiwavelength properties (X-ray - optical - radio). Future work also includes observational follow-up of the small number of optically faint, very low redshift sources identified in this paper. 

\section*{Acknowledgments}

We would like to thank the entire 6dFGS team for providing access to the database prior to the public release. This research made use of data provided by the NASA/IPAC Extragalactic Database (NED) which is operated by the Jet Propulsion Laboratory, California Institute of Technology, under contract with the National Aeronautics and Space Administration. EKM would like to thank R. W. Hunstead and S. Komossa for their helpful comments and assistance. We also thank the referee for his/her useful comments. We acknowledge the support of the Australian Research Council through the award of an ARC Australian Professorial Fellowship to EMS and a QEII Fellowship to SMC.

\bibliographystyle{scemnras}

\bibliography{rasspaper}
  
\bsp

\appendix

\section[Appendix A]{Spectra of sources listed in Table 2.} 

\label{6dfspectraapp}

6dFGS spectra of sources appearing in Table \ref{wrongredshiftstab} where we believe the 6dFGS redshift is more reliable. The dashed lines represent where features would occur at a given redshift and do not neccessarily mean that all the marked features are evident.

\begin{figure*}
  \centerline{\epsfig{file=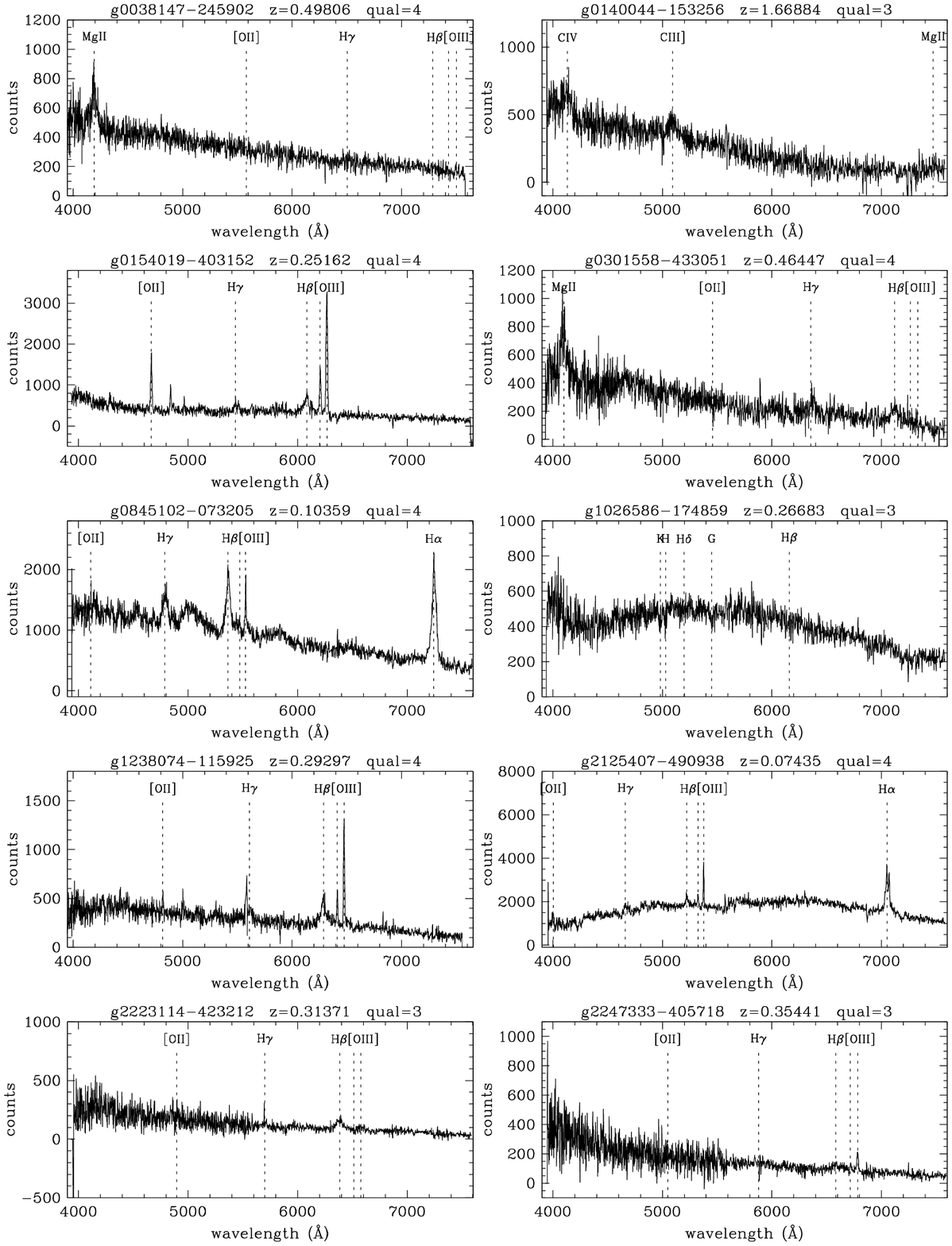, width=\linewidth}}
\end{figure*}

\label{lastpage}

\end{document}